\documentclass[conference]{IEEEtran}
\IEEEoverridecommandlockouts
\ifCLASSINFOpdf
\else
\fi
\usepackage{amsmath,amssymb}
\usepackage{mathtools}
\usepackage{url,hyperref}
\usepackage{graphicx}
\usepackage[dvipsnames]{xcolor}
\definecolor{darkred}{rgb}{0.49,0.04,0.01}
\usepackage{booktabs} 
\usepackage{tabularx}
\usepackage{tcolorbox}

\newcolumntype{L}[1]{>{\raggedright\arraybackslash}p{#1}}
\newcolumntype{C}[1]{>{\centering\arraybackslash}p{#1}}
\newcolumntype{R}[1]{>{\raggedleft\arraybackslash}p{#1}}

\begin{document}
%
\title{An Interpretable Alternative to Neural Representation Learning
for Rating Prediction - Transparent Latent Class Modeling of User Reviews}

\author{\IEEEauthorblockN{Giuseppe Serra\textsuperscript{\textasteriskcentered}}\thanks{\textsuperscript{\textasteriskcentered}Work done when affiliated with NEC Labs Europe and with the University of Birmingham.}
\IEEEauthorblockA{Goethe University Frankfurt \\
Germany \\
\small \texttt{serra@med.uni-frankfurt.de}}
\and
\IEEEauthorblockN{Peter Ti\v{n}o}
\IEEEauthorblockA{University of Birmingham \\
United Kingdom \\
\small \texttt{p.tino@cs.bham.ac.uk}}
\and
\IEEEauthorblockN{Zhao Xu}
\IEEEauthorblockA{NEC Labs Europe \\
Germany \\
\small \texttt{zhao.xu@neclab.eu}}
\and
\IEEEauthorblockN{Xin Yao}
\IEEEauthorblockA{SUSTech\\
China \\
\small \texttt{xiny@sustech.edu.cn}}}

%


\maketitle

\begin{abstract}
Nowadays, neural network (NN) and deep learning (DL) techniques are widely adopted in many applications, including recommender systems. Given the sparse and stochastic nature of collaborative filtering (CF) data, recent works have critically analyzed the effective improvement of neural-based approaches compared to simpler and often transparent algorithms for recommendation. Previous results showed that NN and DL models can be outperformed by traditional algorithms in many tasks. Moreover, given the largely black-box nature of neural-based methods, interpretable results are not naturally obtained. Following on this debate, we first present a transparent probabilistic model that topologically organizes user and product latent classes based on the review information. In contrast to popular neural techniques for representation learning, we readily obtain a statistical, visualization-friendly tool that can be easily inspected to understand user and product characteristics from a textual-based perspective. Then, given the limitations of common embedding techniques, we investigate the possibility of using the estimated interpretable quantities as model input for a rating prediction task. To contribute to the recent debates, we evaluate our results in terms of both capacity for interpretability and predictive performances in comparison with popular text-based neural approaches. The results demonstrate that the proposed latent class representations can yield competitive predictive performances, compared to popular, but difficult-to-interpret approaches.
\end{abstract}


%
\IEEEpeerreviewmaketitle

\section{Introduction} \label{sec:intro}
 In recent years, with the advent of more powerful and capable machines, researchers have started focusing more and more on developing (deep) neural architectures for a wide range of applications. The success of neural-based approaches in different domains, as language modeling \cite{Lample2018phrase,Devlin2019bert} or computer vision \cite{Parmar2018image,Dosovitskiy2020image}, led these models to also dominate the recommender systems research area \cite{Li2017neural,Ying2018graph,Wu2019session}. However, recent works have raised some concerns about the 
relative performance improvements of deep learning approaches compared to simpler algorithms for recommendation tasks \cite{Zhang2020,Dacrema2019,Dacrema2021}. Indeed, as shown in previous works \cite{Lin2019,Ludewig2018}, new recommendation methods do not significantly outperform existing approaches or even can be outperformed by very simple methods, e.g. nearest-neighbor-based techniques \cite{Jannach2017recurrent}. Previous investigations in this direction were mainly focused on pure collaborative filtering (CF) data, where the only available input is the rating matrix. Nevertheless, recent studies in this area tend to conclude that numerical rating data are not informative enough for discovering user preferences. Consequently, given the availability of large collections of textual data, such as product reviews or social media posts, many approaches have tried to extend and improve recommendation models by leveraging such textual information \cite{Mcauley2013,Tu2017,Zheng2017,Tay2018,Garcia2020}. 

In fact, corpora of textual documents contain a wealth of information. 
They can be used to improve the predictive performances of recommendation systems, but more importantly, they provide human understandable explanation about user preference and product properties. To integrate the extra textual information into recommendation systems, most existing works employ embedding methods, i.e., representing items, words, and documents as vectors (also known as embedding) for better flexibility. Although the embedding-based methods often provide good predictions, 
the resulting embeddings are usually not explainable; if singularly evaluated, a 100D or 200D vectors
can be hard to comprehend by humans.

Inspired by recent discussions about the \textit{``neural hype''} \cite{Lin2019}, in this paper we investigate whether using a simpler, more transparent and principled way to learn user and product latent representations can lead to comparable results in the 
recommendation task, i.e. review-based rating prediction.
We present a probabilistic framework for the topographic organization of review data. In contrast with previous neural-based works, we impose a double two-dimensional topological organization of user and product latent classes based on the textual information.
As a result, the latent classes of users and products are organized on two different square grids that reflect the textual input space. The grid organization makes the investigation and interpretation of the results fast and intuitive. Additionally, the probabilistic assumptions of our system enable us to analyze the extracted information in a statistical manner. 

With the interpretable topographically organized latent space representations obtained in our probabilistic framework, one can naturally solve many downstream learning tasks, such as product rating prediction. 
Motivated by the strong correlation between reviews and ratings, we believe that exploiting our simple, yet meaningful representations of textual review information will 
lead to competitive rating prediction results, while maintaining the explainability of the latent classes. We compare our results with state-of-the-art approaches, including both 
neural network (NN) based methods and non-NN ones.
The main goal of this work is not to develop a new approach that outperforms the state-of-the-art with respect to a single predictive score (e.g. accuracy or AUC), but rather to propose a principled and interpretable method that focuses on learning a transparent latent structure of the review data, while having competitive rating prediction performances. Finally, inspired by the recent debates on the \textit{phantom progress} \cite{Dacrema2021} of NN-based methods for recommendation tasks, we compare our findings in terms of both interpretability and predictive performances with respect to the most popular text-based neural approaches for rating prediction.

In summary, the main contributions of this paper are threefold:
\begin{enumerate}
    \item We present a topographic organization of user and product latent classes based on the latent structure of the review data. Common embedding techniques model \textit{each} item by employing dense and complicated representations. In our case, we model \textit{classes} of users and products resulting on latent vectors that are compact, interpretable and rather discrete. Also, different from existing works where the analysis of the user and product characteristics from the textual perspective is practically ignored, we propose a more complete investigation of the available information.
    \item In contrast with previous works, the probabilistic assumptions of the framework allow us to explicitly impose through model formulation the interpretability of the latent codes. Differently from neural-based models where explanations are typically generated through a modification of a difficult-to-interpret network (e.g., by integrating an attention mechanism), in our case, interpretability is at the core of the proposed approach. Rather than implementing an architecture capable of creating explanations without a true understanding of the data, we present a probabilistic model grounded on data understanding which, in turn, is inherently interpretable. In addition, the topographic organization can help us to understand the relationships among different latent classes. Indeed, classes close to each other in the topographic maps should exhibit similar patterns.
    \item Motivated by the strong correlation between reviews and ratings, we exploit our representations of the review information as input for a rating prediction task. We contribute to the debate about the effective performances of neural-based approaches in comparison with more principled and simpler methods. Differently from previous investigations, we present a comprehensive comparison mainly focused on text-based approaches for rating prediction, considering both interpretability capacity and predictive performances.
\end{enumerate}

The rest of the paper is organized as follows. We start off with a brief review of related works. Afterwards we describe the proposed model. Before concluding, we present the experimental results on multiple data sets. In the experimental section, we perform both quantitative and qualitative evaluations of the proposed method.

\section{Related Works} \label{sec:related_works}
There are two lines of research related to the work.

\subsection{Topographic Organization in Latent Models}
Kohonen's seminal work on self-organizing maps (SOM) \cite{Kohonen1982} was introduced in the 1980s. Given the ability to produce low-dimensional representations of high-dimensional data while providing a good approximation of the input space, many extensions and advancements have been proposed in later years. Generative topographic mapping (GTM) \cite{Bishop1998} is one of the most popular probabilistic alternatives to SOM. GTM provides a generative extension of the SOM, assuming a specific discretised non-Gaussian latent prior. Applications and extensions of SOM span many different domains. For example, in \cite{Hofmann2000}, the author proposed a data-driven, statistical approach to visualize large collections of text documents using two-dimensional maps. In CF applications, \cite{Polcicova2004} introduced a topographic organization of latent classes for rating preferences. Differently from their work, where the user preferences are organized using the numerical information, in this paper, we propose to induce a topographic organization of both user and product latent classes exploiting the associated textual information. As a result, we obtain two separate grids (one for users and one for products) reflecting the word patterns of the data. As reported in \cite{Hofmann2000}, the most nuanced and sophisticated medium to express our feelings is our language. 
We believe that it is important to understand and organize the review information in a structured and intuitive way. 

\subsection{Text-based Recommendation Models}
Despite the statistical foundation and the nice visualization capabilities of the previous methods, with the advent of more powerful and capable machines, researchers started focusing more and more on developing deep neural architectures for recommendation. More generally, for this task, one of the most popular approaches is matrix factorization (MF) and, specifically, Singular Value Decomposition (SVD). This method maps users and items into a latent factor space and computes the rating as a dot product between the user and the product embeddings. Although such approaches are effective and simple, the results are poorly interpretable. Indeed, the embeddings of users and items are not explainable and, not knowing what each feature means, it is impossible to unveil user preferences and product characteristics \cite{Seo2017}.

Given the availability of large collections of product reviews, researchers have recently extended latent factor models to leverage the textual information for improving rating prediction performances. In fact, recent studies in this area tend to conclude that the numerical rating information is not powerful enough for discovering user preferences. One of the first attempts that demonstrates the usefulness of leveraging features extracted from reviews to improve the rating prediction accuracy was presented in \cite{Jakob2009}. Among other popular works in this direction, Hidden Factors as Topics (HFT) \cite{Mcauley2013} learns topics using a Latent Dirichlet Allocation (LDA)-like model for each item and an MF for ratings. Ratings Meets Reviews (RMR) \cite{Ling2014} uses the same LDA model for modeling the textual information, but it uses Gaussian mixtures for the rating prediction part. Similarly, TopicMF \cite{Bao2014} learns topics from each review. In \cite{Almahairi2015}, instead of using an approach based on LDA, the explored methods are neural network-based. Most of the existing works combine two learning objectives; one (unsupervised) for the textual information, and one (supervised) for rating prediction. The unsupervised loss acts as a regularization term for the rating prediction loss while taking advantage of the review data. Consequently, the vector representations of the reviews are learned to work well for rating prediction tasks while preventing overfitting.  

Another category of models uses deep learning approaches for learning latent representations of users and items. These methods mainly differ in the neural architecture they use. In \cite{Seo2017}, the authors propose an attention-based CNN (Attn+CNN) to build vector representations of users and products. In \cite{Wang2015} they propose a hierarchical Bayesian model called Collaborative Deep Learning (CDL), which jointly performs deep representation learning and CF for the rating matrix. In \cite{Garcia2020}, the model learns vector representations of users, items, and reviews. The review embedding is learned as a translation in the vector space between the user and the product embeddings.

Even though many works have been published in this direction, as pointed out in \cite{Zhang2020} and \cite{Dacrema2019}, it is debatable whether deep-learning-based models are really making progress in this research area.  
Additionally, the majority of the existing works deal with embedding, but the resulting vector representations usually do not reflect any visualization-driven assumption of the data, making the output poorly explainable. Indeed, if singularly evaluated, common embeddings are not informative. Also, since the ultimate goal is to improve the rating prediction part, the textual information is solely used as an additional source to achieve this goal. Consequently, the latent information contained in the textual data is not fully exploited and investigated. Moreover, there have been related methods in the direction of explainable recommendations. These works address the problem of generating explanations through knowledge graph reasoning \cite{xian2020cafe}, neural attentive models \cite{chen2018neural,yu2019nairs,dong2020asymmetrical}, attraction modeling \cite{hu2018interpretable} or substitute recommendation systems \cite{chen2020try}. However, they do not aim to explicitly generate interpretable vector representations. Instead, the main objective is to generate user-specific explainable recommendations using the complete textual information and highlighting words that are important to explain item-specific ratings. Differently, in our work, we propose a framework for deeper understanding of the data, presenting a two-step approach that starts from the interpretable organization of the textual information to arrive at the rating prediction task. 

\section{Proposed Framework}
\label{sec:framework}
In this section, we present the key ingredients of our framework. 
Based on our observation that
analysis of the textual latent patterns is often not fully covered, we propose an interpretable review-based probabilistic model for rating prediction. First, we describe the estimation of the model parameters. Then, to make our model able to integrate new user data after training, we propose an out-of-sample extension
allowing us to compute the latent class assignments of new users that reviewed products available in our data set. Finally, we present how to exploit the estimated latent space representations as model inputs for a rating prediction task.

\subsection{Topological Organization of the Latent Model}
\label{sec:latent_model}
Consider a collection of users ${\cal U} = \{u_1, u_2, ..., u_N\}$, products
${\cal P} = \{p_1, p_2, ..., p_M\}$ and words
${\cal V} = \{w_1, w_2, ..., w_V\}$.
The data $\cal D$ is a collection of $R$ triples
${\cal D} = \{(u^i, p^i, r^i)\}_{i=1}^R$,
each triple identifying the user $u^i \in {\cal U}$ writing a review $r^i$ on product
$p^i \in {\cal P}$. The review $r^i$ is a multi-set of words from ${\cal V}$,
$r^i = (w^i_1, w^i_2,...,w^i_{S_i})$,
$w^i_j \in {\cal V}$. The latent variables $\mathbf{z}_u \in \{1, \dots, K\}$ and $\mathbf{z}_p \in \{1, \dots, L\}$ represent \textit{abstract} classes of users and products.

Given a review $i$, the probability of sampling a word $w_j \in r^i$ is modeled as:
\begin{eqnarray}
    \label{eq:1}
    P(w_j^i|u^i,p^i) &=&
    \sum_{k=1}^K \sum_{\ell=1}^L 
    \bigg( P(w_j^i|\mathbf{z}_u = k, \mathbf{z}_p = \ell) \big. \nonumber \\
    && \big. P(\mathbf{z}_u = k|u^i) P(\mathbf{z}_p = \ell|p^i) \bigg).
\end{eqnarray}

We impose a grid topology on latent classes via the channel noise methodology \cite{Hofmann2000,Polcicova2004}. Let's assume $\mathbf{y}$ and $\mathbf{z}$ are two different latent classes in our grid. The channel noise for both the product and user latent class grids is defined using the neighborhood function:
\begin{equation}
    \label{eq:neighborhood}
    P(\mathbf{y}|\mathbf{z}) = \frac{\text{exp}\left(\frac{- \|\mathbf{z} - \mathbf{y}\|^2}{2\sigma^2}\right)}
                    {\sum_{\mathbf{y}'}\text{exp}\left(\frac{-\|\mathbf{z} - \mathbf{y}'\|^2}{2\sigma^2}\right)}
\end{equation}
where $\sigma > 0 $ controls the 'concentration' of the transition probabilities among the neighbors of the latent class $\mathbf{y}$.\footnote{Since the channel noise formulation is the same for both the user and product latent grids we do not use subscripts $u$ or $p$ when referring to the latent class.} Overall, when $\mathbf{y}$ and $\mathbf{z}$ are close to each other on the grid, the probability of being corrupted one into another is higher than when they are distant. Additionally, when $\sigma$ is close to 0, then the transition probabilities are more concentrated around $\mathbf{y}$ than for larger values of $\sigma$. 

\noindent
For each user $u \in \mathcal{U}$, the generative process is as follows:
\begin{enumerate}
    \item the latent class assignment $\mathbf{z}_u = k$ is randomly sampled from the user-conditional probability distribution $P(\cdot | u)$ on $\mathbf{z}_u$;
    \item the latent class assignment may be corrupted by the channel noise $P(\mathbf{y}_u|\mathbf{z}_u = k)$, resulting in a (possibly new) class assignment $\mathbf{y}_u = k^{\prime}$. 
\end{enumerate}

For each product $p \in \mathcal{P}$, we can proceed equivalently to get possibly corrupted assignments $\mathbf{y}_p = \ell^{\prime}$ starting from the sampled assignments $\mathbf{z}_p = \ell$.  
\noindent
The topographic organization has a twofold advantage:
1) it enables a model-based visualization tool of the word patterns that can be easily investigated;
2) it regularises the latent class model and hence prevents \textit{overfitting}, while employing a large number of latent classes \cite{Polcicova2004}.

Finally, the model has now the following form:
\begin{equation}
    \label{eq:model1}
    P(\mathbf{y}_u = k^{\prime}|u^i) = \sum_{k=1}^K P(\mathbf{y}_u = k^{\prime}|\mathbf{z}_u = k) 
                        P(\mathbf{z}_u = k|u^i)
\end{equation}
\begin{equation}
    \label{eq:model2}
    P(\mathbf{y}_p = \ell^{\prime}|p^i) = \sum_{\ell=1}^L P(\mathbf{y}_p = \ell^{\prime}|\mathbf{z}_p = \ell) 
                        P(\mathbf{z}_p = \ell|p^i)
\end{equation}
\begin{align}
    \label{eq:model3}
    P(w_j^i|u^i,p^i) =& \sum_{k^\prime=1}^K \sum_{\ell^\prime=1}^L 
    \bigg( P(w_j^i|\mathbf{y}_u = k^\prime, \mathbf{y}_p = \ell^\prime) \big. \nonumber \\
    & \big. P(\mathbf{y}_u = k^\prime|u^i) P(\mathbf{y}_p = \ell^\prime|p^i) \bigg).
\end{align}

In our model the class-conditional word distribution $P(w_j^i|\mathbf{y}_u = k^\prime, \mathbf{y}_p = \ell^\prime)$ is formulated as multinomial.

\subsection{Inference and Learning}
\label{sec:inference}
Assuming independent data items in $\cal D$, the log-likelihood of the model is:
\begin{equation}
    \label{eq:log-lik}
    \mathcal{L}= \sum_{i=1}^R
   \log P(r^i|u^i,p^i).
\end{equation}
We will consider a simple review model assuming independence of the words appearing in the review \textit{i}:
\begin{equation}
    \label{eq:indipendence}
    P(r^i|u^i,p^i) = \prod_{j=1}^{S_i}
   P(w^i_j|u^i,p^i).
\end{equation}
Hence, the log-likelihood reads:
\begin{equation}
    \label{eq:log-lik2}
    \mathcal{L}= \sum_{i=1}^R
   \sum_{j=1}^{S_i} \log P(w^i_j|u^i,p^i).
\end{equation}
Plugging in eqs. (\ref{eq:model1}--\ref{eq:model3}), we obtain:
\begin{align}
    \label{eq:log-lik_final}
\mathcal{L} = \sum_{i=1}^R
   \sum_{j=1}^{S_i} 
   \log
   \bigg[&
 \sum_{k^\prime=1}^K \sum_{\ell^\prime=1}^L 
P(w^i_j|\mathbf{y}_u = k^\prime, 
\mathbf{y}_p = \ell^\prime) 
\nonumber \\
&\sum_{k=1}^K P(\mathbf{y}_u = k^{\prime}|\mathbf{z}_u = k) 
                        P(\mathbf{z}_u = k|u^i) \nonumber\\
& \sum_{\ell=1}^L P(\mathbf{y}_p = \ell^{\prime}|\mathbf{z}_p = \ell) P(\mathbf{z}_p =  \ell|p^i) \bigg].
\end{align}

For training, we use the Expectation-Maximization (EM) algorithm enabling 
maximum likelihood estimation (MLE) in latent variable models.
It iterates the Expectation (E) and Maximization (M) steps until convergence. Detailed derivations of the following equations are presented in Appendix \ref{appA:derivations}.

\subsection{E-step}
\label{sec:e-step}
In the E-step, the algorithm evaluates the current estimates of the model parameters by computing the expected values of the latent variables. We will denote these quantities using $\hat P(\cdot|\cdot)$. Note that we have two levels of hidden variables. First, given the user and the product they reviewed, we do not know which latent classes $\mathbf{z}_u$ and $\mathbf{z}_p$ represented the (\textit{user, product}) couple when writing the review. Second, we know that the underlying latent classes $\mathbf{z}_u$ and $\mathbf{z}_p$ may have been disrupted to latent classes $\mathbf{y}_u$ and $\mathbf{y}_p$ before producing the review, but we do not know their identity.
To simplify mathematical notation, we will denote $\mathbf{z}_u = k$,
$\mathbf{z}_p = \ell$,
$\mathbf{y}_u = k'$ and
$\mathbf{y}_p = \ell'$
as $\mathbf{z}_{u}^k$,
$\mathbf{z}_{p}^\ell$,
$\mathbf{y}_{u}^{k'}$ and
$\mathbf{y}_{p}^{\ell'}$,
respectively.
$\hat P({\mathbf{z}}_{u}^k,{\mathbf{z}}_{p}^\ell|u, p, w)$ is evaluated as:
\begin{equation}
    \frac{P(\mathbf{z}_{u}^k|u) P(\mathbf{z}_{p}^\ell|p)\sum_{k'}\sum_{\ell'}S(k, \ell)}
    {\sum\limits_{k^{\prime\prime}}\sum\limits_{\ell^{\prime\prime}}P(\mathbf{z}_{u}^k|u) P(\mathbf{z}_{p}^\ell|p)\sum\limits_{k'}\sum\limits_{\ell'}S(k^{\prime\prime}, \ell^{\prime\prime})}
\end{equation}
with
\begin{itemize}
\setlength\itemsep{0.2em}
    \item[-] $S(\alpha, \beta) = P(w|\mathbf{y}_{u}^{k^\prime},\mathbf{y}_{p}^{\ell^\prime}) P(\mathbf{y}_{u}^{k^\prime}|\mathbf{z}_{u}^\alpha)P(\mathbf{y}_{p}^{\ell^\prime}|\mathbf{z}_{p}^\beta)$.
\end{itemize}
Instead, $\hat P({\mathbf{y}}_{u}^{k^\prime},{\mathbf{y}}_{p}^{\ell^\prime}|w, u, p)$ is computed as:
\begin{equation}
 \frac{P(w|\mathbf{y}_{u}^{k^\prime},\mathbf{y}_{p}^{\ell^\prime})
    \sum_{k} G(k^\prime)\sum_{\ell} W(\ell^\prime)
    }
    {
    \sum\limits_{k^{\prime\prime}}\sum\limits_{\ell^{\prime\prime}} P(w|\mathbf{y}_{u}^{k^{\prime\prime}},\mathbf{y}_{p}^{\ell^{\prime\prime}})
    \sum\limits_{k} G(k^{\prime\prime})
    \sum\limits_{\ell} W(\ell^{\prime\prime})}
\end{equation}
where 
\begin{itemize}
\setlength\itemsep{0.2em}
    \item[-] $G(\alpha) = P(\mathbf{y}_{u}^{\alpha}| \mathbf{z}_{u}^k)P(\mathbf{z}_{u}^k|u)$
    \item[-] $W(\beta) = P(\mathbf{y}_{p}^{\beta}| \mathbf{z}_{p}^\ell) P(\mathbf{z}_{p}^\ell|p)$.
\end{itemize}

\subsection{M-step}
\label{sec:m-step}
In the M-step, the algorithm maximizes the expectation computed in the E-step by re-estimating the model parameters. To do so, we need to specify functional forms of the distributions for $P(\mathbf{z}_{u}^k|u)$, $P(\mathbf{z}_{p}^\ell|p)$ and $P(w|\mathbf{y}_{u}^{k^\prime},\mathbf{y}_{p}^{\ell^\prime})$. It is natural to model these distributions as multinomial distributions. Thus, we assume:
\begin{align}
    P(\mathbf{z}_{u}^k|u) &\sim \text{Multinomial} \\
    P(\mathbf{z}_{p}^\ell|p) &\sim \text{Multinomial} \\
    P(w|\mathbf{y}_{u}^{k^\prime},\mathbf{y}_{p}^{\ell^\prime}) &\sim \text{Multinomial}.
\end{align}
The update equation for $P(w|\mathbf{y}_{u}^{k^\prime},\mathbf{y}_{p}^{\ell^\prime})$ is:
\begin{equation}
\begin{split}
    P(w|\mathbf{y}_{u}^{k^\prime},\mathbf{y}_{p}^{\ell^\prime}) &= 
    \frac{\sum\limits_{(u,p) \in {\cal B}(w)} \hat P(\mathbf{{y}}_{u}^{k^\prime},\mathbf{{y}}_{p}^{\ell^\prime}| u, p, w)}
        {\sum\limits_{w^\prime} \sum\limits_{(u,p) \in {\cal B}(w^\prime)}  \hat P(\mathbf{{y}}_{u}^{k^\prime},\mathbf{{y}}_{p}^{\ell^\prime}| u, p, w^\prime)}
\end{split}
\end{equation}
where ${\cal B}(w)$ is the set of (\textit{user, product}) tuples associated with the word $w$.

Denoting the set of words used by user $u$ to review product $p$ by ${\cal W}(u,p)$, we obtain the update equations for $P(\mathbf{z}_{u}^k|u)$ and $P(\mathbf{z}_{p}^\ell|p)$ as:
\begin{equation}
\begin{split}
    P(\mathbf{z}_{u}^k|u) &= 
    \frac{\sum_p \sum_{w \in{\cal W}(u,p)} \sum_\ell \hat P(\mathbf{{z}}_{u}^k,\mathbf{{z}}_{p}^{\ell}| u, p, w)}{\sum_p |{\cal W}(u,p)|}
\end{split}
\end{equation}

\begin{equation}
\begin{split}
    P(\mathbf{z}_{p}^\ell|p) &= 
    \frac{\sum_u \sum_{w \in{\cal W}(u,p)} \sum_k 
    \hat P(\mathbf{{z}}_{u}^k,\mathbf{{z}}_{p}^{\ell}| u, p, w)}{\sum_u |{\cal W}(u,p)|},
\end{split}
\end{equation}
where $|{\cal W}(u,p)|$ is the size of ${\cal W}(u,p)$.

\begin{figure}[b]
	\centering
	{\includegraphics[width=.35\textwidth]{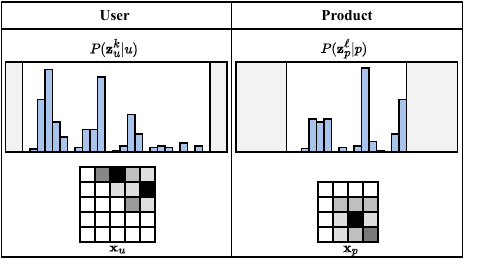}}
	\caption{Input example for our architecture. Each pixel represents the value of the corresponding latent class probability.}
	\label{fig:input_example}
\end{figure}

\subsection{Topographic Initialization with SOM}
\label{sec:som}
The successful application of the EM algorithm depends on the initial position in the parameter space since the algorithm can be sensitive to parameter initialization. We follow an approach similar to the one presented in \cite{Polcicova2004}. Different from their work, where the SOM was run on a dataset of user ratings, we run two different instances of SOM (one for users and one for products) using the review information. We set the number of nodes in SOM to be equivalent to the number of latent classes, i.e. $K$ for users and $L$ for products. We denote by $\mathcal{V}$ the vocabulary set. For each user $u$, there is an associated  $|\mathcal{V}|$-dimensional vector $\mathbf{v}_u$. Each vector dimension represents a word $w$ in the vocabulary and the corresponding value is the term frequency, i.e. how many times the user has written a review using that word. The analogous reasoning is valid for products. In the latter case, each vector dimension represents how many times a word $w$ has been used for reviewing the product $p$. 

After training the two instances of SOM, the conditional latent priors for users and products, i.e. $P(\mathbf{z}_{u}^k|u)$ and $P(\mathbf{z}_{p}^\ell|p)$ respectively, are computed as follows. If the user $u$ belongs to the SOM cluster node $k \in \mathcal{Z}_u=\{1, \dots, K\}$, then we \textit{softened} the hard assignment using the following transformation: 
\begin{equation}
    \label{eq:prior_dist}
    P(\mathbf{z}_{u}^k|u) = 
    \begin{cases}
            A\phantom{(1-A)/(K-1)}  \qquad \text{if } u \in k\\
            (1-A)/(K-1)\phantom{A} \qquad \text{otherwise.}
    \end{cases}
\end{equation}
The parameter $A$ is defined such that if the user $u$ belongs to the class $k$, $P(\mathbf{z}_{u}^k|u)=A$ should be $B>1$ times higher than $P(\mathbf{z}_{u}^{k'}|u)$ for all the other $k' \in \mathcal{Z}_u$. Thus, $A=B/(K-1+B)$. The product conditional latent prior can be generated in an equivalent manner by changing the parameters accordingly. 

The empirical distribution for word patterns $P(w|\mathbf{y}_{u}^{k^\prime},\mathbf{y}_{p}^{\ell^\prime})$ is then computed via the following procedure: (a) to introduce the topology, we follow the steps described in Section \ref{sec:latent_model} to get (possibly corrupted) class indices $\mathbf{y}_u^{k'}$ and $\mathbf{y}_p^{\ell^\prime}$ for all users $u$ and products $p$; (b) we denote with $N(w,k',\ell^\prime)$ the number of times the word $w$ has been used by users belonging to the latent class $k'$ to review products that belong to the latent class $\ell^\prime$. Thus, the empirical distribution is estimated as:
\begin{equation}
    \label{eq:empirical_dist}
    P(w|\mathbf{y}_{u}^{k^\prime},\mathbf{y}_{p}^{\ell^\prime}) = 
    \frac{N(w,k',\ell^\prime) + m}{mV + \sum_{w'\in\mathcal{V}} N(w',k',\ell^\prime)}.
\end{equation}

Due to sparseness, we apply a Laplace correction for smoothing the empirical distribution. The parameter $m$ is a positive number and, in this case, $m=1$. For further details, we refer the reader to \cite{Polcicova2004}.

\subsection{Out-of-sample extension}

\begin{figure}[b]
	\centering
	{\includegraphics[width=.4\textwidth]{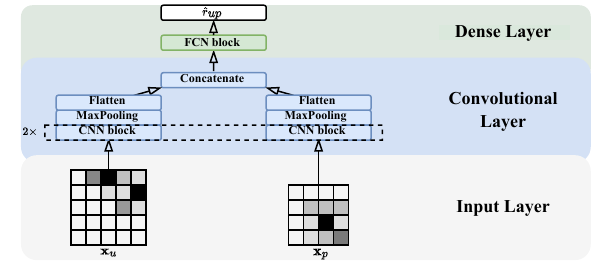}}
	\caption{The schematic view of the proposed architecture. After the input layer, we follow the principles of related CNN-based models for rating prediction tasks.}
	\label{fig:architecture}
\end{figure}

Suppose we have an unknown user $u_{n} \notin \mathcal{U}$ reviewing a product $\bar{p}$ that is available in our dataset. Let's denote with $\bar{r}$ the associated review. Using Bayes' rules, we can extend the model to consider users that were not included during the training phase. In this way, we can learn the latent class assignment of unseen users as follows. 

Using the independence of words assumed in \eqref{eq:indipendence}, we can compute $P(\bar{r}|\mathbf{y}_{u_{n}}^{k^\prime},\bar{p})$ as:
\begin{equation}
    P(\bar{r}|\mathbf{y}_{u_{n}}^{k^\prime},\bar{p}) = \prod_{w \in \bar{r}} P(w|\mathbf{y}_{u_{n}}^{k^\prime},\bar{p})
\end{equation}

where
\begin{equation}
\begin{split}
    P(w|\mathbf{y}_{u_{n}}^{k^\prime},\bar{p}) = \sum_{\ell^\prime=1}^L P(w|\mathbf{y}_{u_{n}}^{k^\prime}, \mathbf{y}_{\bar{p}}^{ \ell^\prime}) P(\mathbf{y}_p^{\ell^\prime}|\bar{p}).
\end{split}
\end{equation}
Consequently, by the Bayes' rule, we have:
\begin{equation}
\begin{split}
    P(\mathbf{y}_{u_{n}}^{k^\prime}|\bar{r}) = \frac{P(\bar{r}|\mathbf{y}_{u_{n}}^{k^\prime},\bar{p}) P(\mathbf{y}_{u_{n}}^{k^\prime}|u_{n})}
    {\sum_{k^{\prime\prime}} P(\bar{r}|\mathbf{y}_{u_{n}}^{k^{\prime\prime}},\bar{p}) P(\mathbf{y}_{u_{n}}^{k^{\prime\prime}}|u_{n})},
\end{split}
\label{eq:out-of-sample}
\end{equation}
assuming $P(\mathbf{y}_{u_{n}}^{k^\cdot}|{u_{n}}) \sim \text{Unif}(0,1)$ as flat prior.

\subsection{Rating prediction part}
\label{sec:rating_part}

After running the EM algorithm, we have the estimates of the conditional distributions $P(\mathbf{z}_{u}^k|u)$ and $P(\mathbf{z}_{p}^\ell|p)$ for all users and products in our dataset. These quantities represent the probability assignments of the items to their respective latent classes. For the next phase of our experiment, we will use the encoded review information as input for the rating prediction task. Intuitively, given the topological organization of the latent classes on two-dimensional grids, we can think of these quantities as images where each pixel represents a latent class and the corresponding value is the latent class probability assignment. An example of the model input transformation for a sampled review is given in Fig. \ref{fig:input_example}.
Given the large employment of embedding techniques for representation learning in recommendation systems, we propose to use the learned latent class assignments as representations of users and products instead. In this way, we can generate a self-explainable latent representation of the items that, at the same time, should perform well on a rating prediction task. The schematic view of our architecture for rating prediction is presented in Fig. \ref{fig:architecture}. After the input layer, the CNN-block consists of the common layers used in CNN-based models. We perform two convolution operations with batch normalization and Rectified Linear Units (ReLU) \cite{Agarap2018relu} as activation function. As a result, the training is faster and more stable. Then, we apply a max-pooling operation to take the maximum value of each feature map obtained by the convolution operations. After the max-pooling operation, the convolutional results are reduced to a fixed size vector for both users and products. We concatenate the results given by the max pooling operation, and we use the resulting vector as input for the fully connected network (FCN) block. This block simply represents a neural network with several dense layers. More details on the structure of this block are given in Section \ref{sec:experimental_settings}. The proposed architecture follows some principles of CNN-based models for rating prediction, e.g. \cite{Zheng2017,Seo2017}. We are not interested in getting better results than state-of-the-art approaches, instead, our ultimate goal in this part is to investigate whether our proposed model input is informative enough to get competitive results compared with more specialized techniques and architectures. In theory, one might potentially change the architecture as desired.

The objective is the error of the predictions. Following previous works, this will be validated according to the Mean Squared Error (MSE):
\begin{equation}
    \text{MSE} = \frac{1}{|\mathcal{T}|} \sum_{(u,p) \in \mathcal{T}} (\widehat{\texttt{score}}_{u,p} - \texttt{score}_{u,p})^2
    \label{eq:mse_1}
\end{equation}
where $\mathcal{T}$ represents either the test or the validation set.

\section{Experiments}
\label{sec:experiments}
In this section, we evaluate the performance of our proposed framework.\footnote{\url{https://github.com/GiuseppeSerra93/TLCM}} First, we analyze the probabilistic latent model and the generated organization of user and product latent classes. Second, we investigate the generative extension of our approach. Last, we compare the performance of our model with state-of-the-art approaches for the rating prediction task.

\begin{table}[t]
  	\caption{Statistics of the Preprocessed Data Sets.}
 	\begin{center}
 	\small
 	\begin{tabular}{lrrr}
 	\hline
 	 & Reviews & Users & Items \\
 	\hline
 	Amazon Instant Videos           & 36241       & 5127       & 1685\\
 	Automotive                      & 20285        & 2926       & 1835\\
 	Baby                        	& 156335        & 19442       & 7050\\
 	Beauty                        	& 194130        & 22356       & 12101\\
 	Cell Phones and Acc.            & 191336        & 27864       & 10429\\
 	Digital Music              	    & 64260        & 5539       & 3568\\
 	Grocery and Gourmet Food		& 148902        & 14675       & 8713\\
 	Health and Personal Care	    & 333274        & 38583       & 18534\\
 	Musical Instruments             & 10218        & 1427       & 900\\
 	Office Products                 & 52988        & 4902       & 2420\\
 	Patio, Lawn and Garden          & 13213        & 1684       & 962\\
 	Pet Supplies                    & 152367        & 19848       & 8510\\
 	Sports and Outdoors          	& 289181        & 35588       & 18357\\
 	Tools and Home Improv.          & 133445        & 16634       & 10217\\
 	Toys and Games              	& 161603        & 19405       & 11924\\
  	Videogames                     	& 227859        & 24291       & 10672\\
	\hline
 	\end{tabular}
 	\end{center}
 	\label{tab:statistics_2}
 \end{table}
 
\subsection{Data Sets Description}
\label{sec:data_description}
We evaluate our method with a benchmark dataset commonly used for product rating prediction: \textit{Amazon Product Data}.\footnote{\url{http://jmcauley.ucsd.edu/data/amazon}} The dataset is a collection of product reviews and metadata, divided per category, retrieved from May 1999 to July 2014. We focus on the \textit{5-core} version of the data sets, where each user and item has at least 5 associated reviews. The ratings are integer values between 1 and 5. We use data from 16 categories for our experiments. Following \cite{Serra2021interpreting}, the review text has been normalized by: (a) setting the maximum length of a raw review to 300; (b) lowering case letters; (c) removing stopwords, numbers, and special characters; (d) removing non-existing words using an English vocabulary as a filter; (e) lemmatization; (f) removing reviews with just one word.
Since the textual information shows strong diversity depending on the product category, we treat each dataset independently. Hence, we select a vocabulary for each category. The vocabulary size is $V=2000$ for all product categories. The word selection is based on a modified version of the \textit{tf-idf} index. After the vocabulary selection, we further filter the reviews to remove the ones without any word belonging to the corresponding vocabulary. Some statistics of the preprocessed data sets are summarized in Table \ref{tab:statistics_2}. From the table, we can observe that the sample size of the data may vary consistently depending on the category. For example, the largest category (\textit{Beauty}) has around 30 times as many reviews as the smallest one (\textit{Musical instruments}). Similarly, the number of users and products oscillates within a wide range of values among different categories. The data sets represent different and realistic conditions; this acts as a good framework for evaluating the quality of the results.

\subsection{Experimental Settings}
\label{sec:experimental_settings}
We set the number of user latent classes $K=25$ and product classes $L=16$; we evaluated the robustness of our method to changes in the hyperparameters $K$ and $L$ but did not observe any significant difference in both qualitative and quantitative analysis. This is in line with the constraint imposed by the topological organization; using a larger number of latent classes did not lead to overfitting. The \textit{specificity} parameters for users and products, defined in \eqref{eq:neighborhood}, are set respectively to $\sigma_u=3$ and $\sigma_p=2$. For the SOM initialization (Section \ref{sec:som}), we used the Python package \textsc{Minisom}.\footnote{\url{https://github.com/JustGlowing/minisom}}

For the probabilistic model part, following previous works (e.g. \cite{Polcicova2004, Hofmann2001}), we apply the so-called \textit{all but one protocol}. From each user having at least $10$ reviews, we randomly select one review to be assigned to the test set. The main focus of this part of the experiment is on discovering word patterns of users and products within large collections of review data, hence we do not need a generative formulation of the model. We evaluate the training procedure according to the normalized negative log-likelihood (NLL). Based on \eqref{eq:log-lik}, we have:
\begin{equation}
    NLL_{\mathcal{T}} = -\frac{1}{|\mathcal{T}|} \sum_{(u,p,r) \in \mathcal{T}} \log \hat P(r|u,p)
\end{equation}
where $\mathcal{T}$ represents either the training or test set.

For the rating prediction task, as in previous works, the data are randomly split by reviews into training (80\%), validation (10\%), and test (10\%) sets. Additionally, we remove reviews from the validation and test sets if either the associated user or product does not belong to the training data set. The FCN block depicted in Fig. \ref{fig:architecture} is a neural network with four hidden fully-connected layers $[ 128, 64, 32, 16]$. We evaluate different configurations of this block by changing the number of layers and the number of units for each layer. We select the best configuration based on the validation scores. The final experiment was run for $200$ learning iterations and validated every iteration. A single epoch performs \textit{Adam} optimizer \cite{Kingma2014} with a learning rate set to $0.02$ and batch size of $256$. The training metric is the MSE, as defined in \eqref{eq:mse_1}. To prevent overfitting, we monitor the validation set score using early stopping with patience set to 10 epochs. 

\subsection{Topographic Organization of Latent Classes}
\label{sec:topographic_organization_results}
Given the model parameters estimated in Section \ref{sec:m-step}, we can analyze the organization of user and product latent classes by inspecting the associated word distributions. The word distribution for each user latent class, under the assumption of uninformative flat prior over latent classes, is computed as:

\begin{equation}
\begin{split}
    P(w|\mathbf{y}_{u}^{k^\prime}) &=
    \sum_{\ell^\prime=1}^L P(w|\mathbf{y}_{u}^{k^\prime},\mathbf{y}_{p}^{\ell^\prime})
    P(\mathbf{y}_{p}^{\ell^\prime}) \\
    & = \frac{1}{L} \sum_{\ell^\prime=1}^L P(w|\mathbf{y}_{u}^{k^\prime},\mathbf{y}_{p}^{\ell^\prime}).
    \label{eq:distribution_user_classes}
\end{split}
\end{equation}


Analogically, the word-distribution for each product latent class can be derived as follows:

\begin{equation}
\begin{split}
    P(w|\mathbf{y}_{p}^{\ell^\prime}) &=
    \sum_{k^\prime=1}^K P(w|\mathbf{y}_{u}^{k^\prime},\mathbf{y}_{p}^{\ell^\prime})
    P(\mathbf{y}_{u}^{k^\prime}) \\
    & = \frac{1}{K} \sum_{k^\prime=1}^K P(w|\mathbf{y}_{u}^{k^\prime},\mathbf{y}_{p}^{\ell^\prime}).
    \label{eq:distribution_product_classes}
\end{split}
\end{equation}

The visualization of the most probable words associated with each latent class helps us to evaluate the results. The list of the $10$ most probable words for product latent classes of the \textit{Automotive} category is provided in Table \ref{tab:product_word_dist}.

\begin{table}[t]
\caption{Product-class Word Distributions - Automotive Category.}
\centering\tiny\renewcommand{\arraystretch}{0.8} 
\begin{tabular}{|C{1.cm}|C{1.cm}|C{1.cm}|C{1.cm}|}
\hline
 light        & light      & blade   & wax      \\
 bulb         & fit        & wiper   & kit      \\
 bright       & lead       & great   & product  \\
 stock        & great      & fit     & great    \\
 white        & good       & good    & wiper    \\
 brighter     & bright     & product & water    \\
 fit          & price      & well    & snow     \\
 price        & well       & rain    & blade    \\
 lead         & order      & brand   & wipe     \\
 replacement  & horn       & jeep    & rain     \\
\hline
 filter       & great      & product & product  \\
 oil          & good       & hose    & paint    \\
 change       & fit        & great   & great    \\
 fit          & tool       & good    & pad      \\
 price        & price      & tank    & clay     \\
 fuel         & product    & fit     & good     \\
 socket       & code       & well    & wax      \\
 mile         & well       & try     & polish   \\
 wrench       & purchase   & say     & try      \\
 good         & say        & quality & water    \\
\hline
 oil          & good       & product & product  \\
 change       & great      & great   & jack     \\
 fluid        & product    & good    & great    \\
 pump         & mount      & cover   & wash     \\
 drain        & item       & strap   & shine    \\
 price        & quality    & lock    & spray    \\
 leak         & fit        & trailer & trailer  \\
 transmission & price      & hitch   & good     \\
 good         & door       & door    & lift     \\
 great        & hitch      & step    & wax      \\
 \hline
 battery      & plug       & wheel   & towel    \\
 charge       & tender     & leather & wash     \\
 power        & battery    & brush   & water    \\
 charger      & great      & cleaner & gauge    \\
 plug         & product    & seat    & pressure \\
 device       & cable      & great   & cloth    \\
 motorcycle   & motorcycle & level   & cleaning \\
 phone        & good       & product & accurate \\
 compressor   & spark      & trailer & valve    \\
 cord         & winter     & good    & great    \\
\hline
\end{tabular}
\label{tab:product_word_dist}
\end{table}

From the results, we can notice clear patterns of the most probable words associated with each latent class. Additionally, we can observe the topological organization of the latent classes: words patterns are similar for adjacent classes in the grid. For example, adjacent classes at the top left of the grid refer to \textit{lighting accessories}, while going down to the bottom left we can deduce that the latent classes refer to \textit{electrical} and \textit{oil system tools}. On the right part of the grid, instead, we have latent classes referring to \textit{cleaning accessories}. Knowing the probabilistic assignments for a sampled item, we can easily identify the most probable latent classes and evaluate the corresponding word distributions. This visualization tool lends itself well to other types of result investigation. For example, if we are interested in exploring a specific user latent class, one may analyze how the product latent classes change when conditioned on the user class of interest. Theoretically, the product latent classes should change according to the main \textit{topic} of the user latent class, i.e. most of the product latent classes should now be associated with the user class of interest. The same reasoning is valid starting from a specific product latent class. Given these considerations, we provide a flexible and straightforward tool to investigate the latent characteristics of the items, making the results clear and understandable.

\subsection{Out-of-sample Extension Results}
\label{sec:out-of-sample_results}
After evaluating the quality of the latent class organizations, we investigate whether our model is able to correctly assign unseen users to the corresponding latent classes. Given a product category, we randomly select a review $\bar{r}$ written by an unknown user $u_{n}$ from the out-of-sample set. Then, given the review $\bar{r}$, we compute 
$P(\mathbf{y}_{u_{n}}^{k^\prime}|\bar{r})$ as described in \eqref{eq:out-of-sample}.
\begin{figure}[ht]
	\centering
	{\includegraphics[width=.42\textwidth]{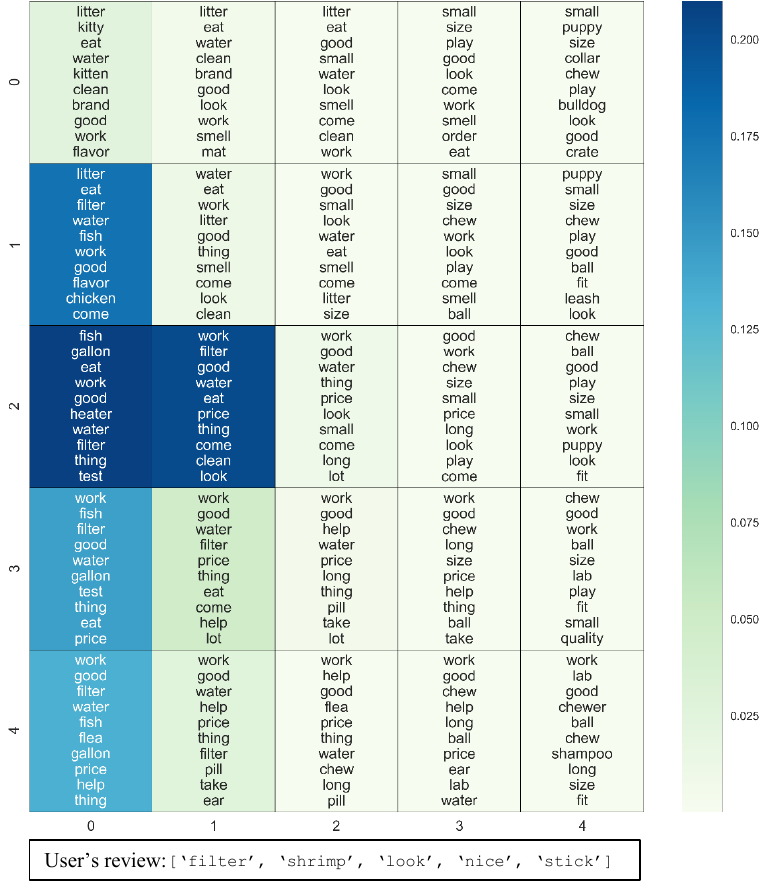}}
	\caption{Results of the out-of-sample extension from the category \textit{Pet Supplies}. The words in the bottom box are the ones used by the unknown user $u_{n}$ to review a product $p$ in our dataset. The considered words are included in the corresponding vocabulary $\mathcal{V}_{\text{pet}}$. Note that the darker the color, the higher is the probability assignment to the corresponding latent class. }
	\label{fig:generative_example}
\end{figure}

Fig. \ref{fig:generative_example} shows the results taken from the \textit{Pet Supplies} category. To make the visualization easier and intuitive, we visualize the learned user latent probability assignments over the user-class word distributions. First, by examining the most probable words associated with each user latent class, we can observe how the latent classes are topologically well organized. Then, by analyzing the words used by the unknown user listed in the bottom box, we can assume that the reviewed product is related to the \textit{aquariums} subcategory. The results show the ability to assign the unknown user to the correct user latent classes. In fact, the most probable latent classes associated with the new user are the ones related to \textit{aquariums}, as easily perceivable by inspecting the corresponding word lists in the grid. It is important to mention that we are able to get meaningful latent class assignments when the words used by the unknown users are, to some extent, informative. For example, if the words were only adjectives (e.g. good, nice) or verbs (e.g. look, work) it would be difficult to classify unknown users correctly. This highlights the importance of selecting a good vocabulary set as a starting point for the evaluation. 

\begin{table*}[ht]
	\caption{MSE for TLCM and State-of-the-art Approaches.}
	\begin{center}
    \begin{small}
	\begin{tabular}{lcc|cccc|cc}
	\hline
	Category & NMF    & SVD       & TNET     & MPCN      & TransRev   & DeepCoNN  & TLCM-CNN & TLCM-LR \\
	\hline
	Amazon Instant Videos  		& 1.628		& 1.206		& 1.007		& 0.997		& \textbf{0.884}		& 0.957		& 0.961		& 1.004 \\
	Automotive  				& 1.171		& 0.818		& 0.946		& 0.861		& 0.855		& \textbf{0.792}		& 0.879		& 0.883 \\
	Baby  						& 2.066		& 1.445		& 1.338		& 1.304		& 1.100		& 1.440		& \textbf{1.094}		& 1.309 \\
	Beauty  					& 2.163		& 1.521		& 1.404		& 1.387		& \textbf{1.158}		& 1.214		& 1.194		& 1.352 \\
	Cell Phones and Acc.		& 2.438		& 1.664		& 1.431		& 1.413		& \textbf{1.279}		& 1.378		& 1.294		& 1.447 \\
	Digital Music 				& 1.787		& 1.381		& 1.004		& 0.970		& 0.\textbf{}782		& 0.808		& 0.836		& 1.070 \\
	Grocery, Gourmet Food  		& 1.879		& 1.370		& 1.129		& 1.125		& \textbf{0.957}		& 1.542		& 1.064		& 1.117 \\
	Health				  		& 2.045		& 1.452		& 1.249		& 1.238		& \textbf{1.011}		& 1.093		& 1.091		& 1.232 \\
	Musical Instruments  		& 1.287		& 0.703		& 1.100		& 0.923		& 0.690		& 0.896		& \textbf{0.670}		& 0.697 \\
	Office Products  			& 0.988		& 0.719		& 0.840		& 0.779		& 0.724		& \textbf{0.723}		& 0.735		& 0.834 \\
	Patio, Lawn and Garden  	& 1.429		& 0.967		& 1.123		& 1.011		& \textbf{0.941}		& 1.020		& 1.045		& 1.100 \\
	Pet Supplies  				& 2.071		& 1.438		& 1.346		& 1.328		& \textbf{1.191}		& 1.447		& 1.244		& 1.381 \\
	Sports			  			& 1.490		& 1.011		& 0.994		& 0.980		& \textbf{0.823}		& 0.898		& 0.898		& 0.958 \\
	Tools and Home Improv.  	& 1.793		& 1.222		& 1.122		& 1.096		& \textbf{0.879}		& 1.208		& 0.954		& 1.091 \\
	Toys and Games  			& 1.542		& 1.111		& 0.974		& 0.973		& \textbf{0.784}		& 0.806		& 0.859		& 0.928 \\
	Videogames  				& 2.188		& 1.629		& 1.276		& 1.257		& \textbf{1.082}		& 1.135		& 1.111		& 1.370 \\
	\hline
	  				& 1.748*		& 1.229*		& 1.143*		& 1.103*		& 0.946*		& 1.085*		& 0.996*		& 1.111* \\
	\hline
	\end{tabular}
	\end{small}
	\end{center}
    \label{tab:MSE}
\end{table*}

\subsection{Rating Prediction Task}
\label{sec:rating_prediction_results}
We compare our method with several baselines considering both factorization-based approaches and methods that take advantage of the textual information for improving the rating prediction performance.

\noindent
    \textbf{Non-negative Matrix Factorization (NMF)} is a standard baseline for CF. The rating prediction is set as the dot product between item and user factors, i.e. $q_i^\top p_u$. The latent factors are kept positive. \\
    \textbf{Singular Value Decomposition (SVD)} is very similar to NMF. The rating is computed as a dot product between the user and the product latent factors. In both cases, the results are poorly explainable. Indeed, it is not possible to understand user and product characteristics from the evaluation of the latent factors.\\
    \textbf{Multi-Point Co-Attention Networks (MPCN)} \cite{Tay2018} proposes a pointer-based model to extract important reviews from user and item reviews, based on the intuition that not all the reviews are equally informative. Once the important reviews are selected, a co-attentive layer learns the most important words associated. In this way, the model can learn the most informative user and product reviews for each user-item pair.\\
    \textbf{Deep Co-Operative Neural Networks (DeepCoNN)} \cite{Zheng2017} learns convolutional user and item representations from the textual information. Even though is a text-based model, the analysis is focused only on the rating prediction performances. Additionally, given the \textit{deep} nature of the representations, the results are poorly interpretable.\\
    \textbf{TransNets (TNET)} \cite{Tu2017} is an extension of the DeepCoNN model. It introduces an additional \textit{transform} layer that learns an approximation of the review corresponding to the target user-item pair and, during the training phase, enforces it to be similar to the embedding of the actual target review. The model can be used to suggest reviews that are more similar to the one potentially written by the target user. The qualitative evaluation just rely on the visualization of some sampled reviews by highlighting the most similar sentences between the original review and the predicted most helpful one. However, if singularly evaluated, the learned embeddings do not provide any interpretable information.\\
    \textbf{TransRev} \cite{Garcia2020} is similar to TransNets in that it learns review embeddings as a translation of the reviewer representation to the reviewed product embedding. Ratings are predicted based on an approximation of the review embedding at test time based on the difference between the embedding of the user and the item. As in TransNets, the approximation is also used to suggest a tentative review to users for further elaborations. The evaluation of the learned word embedding is performed using t-SNE. Items and users are related to reviews by means of non-interpretable latent representations.\\
    \textbf{TLCM-CNN} stands for \textit{Transparent Latent Class Model with CNN} and represents the rating prediction model described in Section \ref{sec:rating_part}. This comparison helps us to investigate if the latent representations of our probabilistic model, even though not learned \textit{ad-hoc}, represent an informative input for the rating prediction task. For the reasons explained before, our input representations, if able to get competitive results, would be better in terms of interpretability capacity compared to the embedding techniques presented in the previous works.\\
    \textbf{TLCM-LR} performs a simple linear regression (LR) that, for each review, takes as input the concatenation of the estimates of the corresponding user and product learned by our probabilistic model. As stated in section \ref{sec:rating_part}, the architecture for rating prediction can be changed as desired. In this case, we decided to use a linear regression model since, among other methods, it is known to be a \textit{transparent} model. Following the direction of previous investigations on the \textit{neural hype}, this experiment allows us to understand whether the use of complicated architectures is helping to make significant progress for recommendation tasks.

As anticipated, the ultimate goal of this work is to build a review-based model suitable for both visualizing and learning user and product characteristics. Then, driven by the limitations of existing approaches to represent users and products through explainable latent vectors, we propose to exploit our estimated quantities as input for rating prediction. This idea is fully motivated by the strong correlation between ratings and reviews. In addition, we also propose to use the same information as input for an interpretable model, i.e. a linear regression, to understand its predictive performances compared to neural approaches. This experimental part completes our analysis about the \textit{``phantom progress"} of neural-based approaches for representation learning and recommendation tasks, considering also the interpretability capacity of the models taken into consideration.

In terms of interpretability, as already mentioned, all the baselines can provide meaningful representations in the latent space, but neither the single numbers contained in the vectors nor their dimensions have an interpretable meaning \cite{Senel2018}. The main limitation of these techniques laid in the fact that they can capture relations among items by using vectors that are only meaningful to each other. For instance, if we try to evaluate user vectors without employing the review vectors, we would not be able to comprehend the latent textual information associated with them. Differently, our model can directly encode the textual information within the estimated quantities, providing vectors that are \textit{self-explainable}. Additionally, our latent representations are more compact and sparse, making them easier to evaluate. Finally, the intuitive visualization properties provided by our square grids create a simple tool to investigate the results, further enhancing their interpretability.

In terms of rating prediction performance, we independently repeat each experiment on five different random splits closely following the experimental setup specified in the relevant studies, including reporting the averaged Mean Squared Error (MSE) as the performance measure. For a fair comparison, we conduct and compare the baselines on the same data splits, when possible, using the default configurations provided by the authors. If not, whether for difficulty in reusing the source code or its unavailability, we directly copy the results from the original papers. In detail, for SVD and NMF we used the Python package \textsc{Surprise}\footnote{\url{http://surpriselib.com/}} and we selected the best learning parameters through grid search. For DeepCoNN, since the original authors' source code has not been released, we used a third-party implementation.\footnote{\url{https://github.com/chenchongthu/DeepCoNN}} In this case, we applied the default hyperparameter setting. For the remaining baselines, we copied the results from the corresponding original papers. Results in terms of MSE are reported in Table \ref{tab:MSE}. The asterisk indicates the macro MSE across all the product categories. From the results, in line with the conclusions of existing works, review-based methods outperform the ones based only on the numerical information. We can observe that our model is able to outperform most of the state-of-the-art approaches in the analyzed categories and has comparable performances with the best one, i.e. TransRev. Additionally, it is worth to mention that our simplest version of the architecture, i.e. TLCM-LR, is also able to get comparable results against most of the state-of-the-art approaches. This demonstrates that: 1) by carefully encoding the textual information into the latent representations, it is possible to get similar results in comparison with more complicated techniques that provide dense and high-dimensional vector representations of the items; b) in line with the conclusions of previous investigations on CF data, also in case the textual information is used, we do not need to employ complicated neural-based models to get competitive results. The latent representations proposed in our work, even though not learned to specifically perform well on a rating prediction task, are able to maintain competitiveness in comparison with more specialized architectures. This clearly suggests that one may potentially use the proposed model input with nice interpretable properties, without worsening the rating prediction task considerably.

\section{Conclusion}
\label{sec:conclusions}
Recently, several approaches for rating predictions of textual reviews in the framework of deep neural networks have appeared in the literature \cite{Mcauley2013,Almahairi2015,Tu2017,Zheng2017,Tay2018,Garcia2020}. Given the highly stochastic nature of the data and relative data sparsity, one can legitimately ask to what extent can the full predictive power of deep networks be utilized in this context. The question is even more relevant when one realises that clear interpretability of the deep network functionality is still an open problem \cite{Rudin2018stop}. To answer this question, we present an approach for product rating prediction using a relatively simple and interpretable latent class probabilistic model utilizing topographic organization of user and product latent classes based on the review information. In existing works, the review information is usually exploited to enhance the rating prediction performances, but is not fully inspected to understand user and product features. In contrast, we propose a deeper understanding of the data, presenting a two-step approach that starts from the interpretable organization of textual information to arrive at the rating prediction task. The organization of the latent classes on 2-dimensional grids provides a visualization tool that can be used to statistically investigate user and product features from a review-based perspective. Through this organization, we can arrange complicated and unstructured textual data in a simple way. The thorough analysis in the experimental section demonstrates the ease of analyzing the latent review patterns using tools from probabilistic theory. The visualization of the results, presented in sections \ref{sec:topographic_organization_results} and \ref{sec:out-of-sample_results}, shows that the lower-dimensional latent representations of users and products are a good approximation of the textual input space. Consequently, driven by the assumption that ratings and reviews are strongly correlated, we propose to use the resulting latent features as input for a rating prediction task. In this part, we contribute to the debate about the \textit{``phantom progress"} of deep learning approaches for recommendation tasks. Our investigation, differently from the previous ones, is mainly focused on methods that take advantage of the textual information for rating prediction tasks, and includes an evaluation that considers the capacity of such models to represent users, products, and reviews utilizing interpretable latent vectors. The results suggest that, also in the textual-based case, the use of dense and complicated representations is not fully motivated. Indeed, even though our representations are not learned for a rating prediction task specifically, the results are comparable to models that learn \textit{ad-hoc} representations. Nonetheless, being highly interpretable, the proposed latent representations overcome the limitations of the common embedding techniques used in most of the considered previous works. Finally, the prediction results of our linear regression model suggest that we do not need to implement deep architectures either. This simple model is able to outperform some of the baselines and get comparable results with the remaining ones, while being fully transparent. Naturally, there is always a trade-off between model capabilities and interpretability. The nice and explainable visualization properties of our constrained model may affect the modeling capabilities for rating prediction. On the other hand, better modeling capabilities through common embedding techniques and deep architectures provide representations that are created for performing well on a specific task, but at the price of losing the human interpretation of the results.


\section*{Acknowledgment}
This  project  has  received  funding  from  the  European Union’s  Horizon  2020  research  and  innovation  programme under grant agreement No 766186.

\bibliographystyle{IEEEtran}
\bibliography{reference}

\appendices
\section{Derivations of EM Equations}
\label{appA:derivations}
\subsection{E-step}
First, we estimate $\hat P({\mathbf{z}}_{u}^k, {\mathbf{z}}_{p}^\ell|w, u, p)$ as:
\begin{equation}
    \frac{P(w|u, p, \mathbf{z}_{u}^k, \mathbf{z}_{p}^\ell) P(\mathbf{z}_{u}^k|u, p)P(\mathbf{z}_{p}^\ell|u, p)}
    {\sum\limits_{k^{\prime\prime}} \sum\limits_{\ell^{\prime\prime}} P(w|u, p, \mathbf{z}_u^{k^{\prime\prime}},\mathbf{z}_p^{\ell^{\prime\prime}}) P(\mathbf{z}_u^{k^{\prime\prime}}|u, p)P(\mathbf{z}_p^{\ell^{\prime\prime}}|u, p)}.
\end{equation}
By model assumptions, $\mathbf{z}_u$ is independent from products and $\mathbf{z}_p$ is independent from users, so we have $P(\mathbf{z}_{u}^k|u, p) = P(\mathbf{z}_{u}^k|u)$ and $P(\mathbf{z}_{p}^\ell|u, p) = P(\mathbf{z}_{p}^\ell|p)$. Consequently:

\begin{equation}
\begin{split}
    P(w|&u, p, \mathbf{z}_{u}^k, \mathbf{z}_{p}^\ell) = \sum_{k^\prime} \sum_{\ell^\prime} P(w, \mathbf{y}_{u}^{k^\prime},\mathbf{y}_{p}^{\ell^\prime}|u, p,\mathbf{z}_{u}^k,\mathbf{z}_{p}^\ell)
    \\
    &= \sum_{k^\prime} \sum_{\ell^\prime} \bigg[
    P(w| u, p,\mathbf{z}_{u}^k,\mathbf{z}_{p}^\ell,\mathbf{y}_{u}^{k^\prime},\mathbf{y}_{p}^{\ell^\prime}) \\
    &
    \qquad P(\mathbf{y}_{u}^{k^\prime}|u, p, \mathbf{z}_{u}^k)
    P(\mathbf{y}_{p}^{\ell^\prime}|u, p, \mathbf{z}_{p}^\ell) \bigg]
    \\
    &= \sum_{k^\prime} \sum_{\ell^\prime} P(w| \mathbf{y}_{u}^{k^\prime},\mathbf{y}_{p}^{\ell^\prime}) P(\mathbf{y}_{u}^{k^\prime}|\mathbf{z}_{u}^k)P(\mathbf{y}_{p}^{\ell^\prime}|\mathbf{z}_{p}^\ell),
\end{split}
\end{equation}
and so $\hat P({\mathbf{z}}_{u}^k,{\mathbf{z}}_{p}^\ell|w, u, p) $ is equal to

\begin{equation}
    \frac{P(\mathbf{z}_{u}^k|u) P(\mathbf{z}_{p}^\ell|p)\sum_{k'}\sum_{\ell'}S(k, \ell)}
    {\sum\limits_{k^{\prime\prime}}\sum\limits_{\ell^{\prime\prime}}P(\mathbf{z}_{u}^k|u) P(\mathbf{z}_{p}^\ell|p)\sum\limits_{k'}\sum\limits_{\ell'}S(k^{\prime\prime}, \ell^{\prime\prime})},
\end{equation}
with
\begin{itemize}
\setlength\itemsep{0.2em}
    \item[-] $S(\alpha, \beta) = P(w|\mathbf{y}_{u}^{k^\prime},\mathbf{y}_{p}^{\ell^\prime}) P(\mathbf{y}_{u}^{k^\prime}|\mathbf{z}_{u}^\alpha)P(\mathbf{y}_{p}^{\ell^\prime}|\mathbf{z}_{p}^\beta)$.
\end{itemize}

Analogously, to compute $\hat P({\mathbf{y}}_{u}^{k^\prime},{\mathbf{y}}_{p}^{\ell^\prime}|w, u, p)$, we have:
\begin{equation}
    \frac{P(w|u, p, \mathbf{y}_{u}^{k^\prime},\mathbf{y}_{p}^{\ell^\prime})
    P(\mathbf{y}_{u}^{k^\prime}|u, p)P(\mathbf{y}_{p}^{\ell^\prime}|u, p)
    }
    {
    \sum_{k^{\prime\prime}}\sum_{\ell^{\prime\prime}}
    P(w|u, p, \mathbf{y}_{u}^{k^{\prime\prime}},\mathbf{y}_{p}^{\ell^{\prime\prime}})
    P(\mathbf{y}_{u}^{k^{\prime\prime}}|u, p)P(\mathbf{y}_{p}^{\ell^{\prime\prime}}|u, p)}
\end{equation}
with
\begin{equation}
\begin{split}
    P(\mathbf{y}_{u}^{k^\prime}|u, p)  &= \sum_{k} P(\mathbf{y}_{u}^{k^\prime}, \mathbf{z}_{u}^k|u, p) \\
        &= \sum_{k} P(\mathbf{y}_{u}^{k^\prime}| u, p, \mathbf{z}_{u}^k) P(\mathbf{z}_{u}^k|u, p) \\
        &= \sum_{k} P(\mathbf{y}_{u}^{k^\prime}| \mathbf{z}_{u}^k) P(\mathbf{z}_{u}^k|u).
\end{split}
\end{equation}
Following the same reasoning:
\begin{equation}
\begin{split}
    P(\mathbf{y}_{p}^{\ell^\prime}|u, p)  &= \sum_{\ell} P(\mathbf{y}_{p}^{\ell^\prime}, \mathbf{z}_{p}^\ell|u, p) \\
        &= \sum_{\ell} P(\mathbf{y}_{p}^{\ell^\prime}| u, p, \mathbf{z}_{p}^\ell) P(\mathbf{z}_{p}^\ell|u, p) \\
        &= \sum_{\ell} P(\mathbf{y}_{p}^{\ell^\prime}| \mathbf{z}_{p}^\ell) P(\mathbf{z}_{p}^\ell|p).
\end{split}
\end{equation}
Hence, $\hat P({\mathbf{y}}_{u}^{k^\prime},{\mathbf{y}}_{p}^{\ell^\prime}|w, u, p)$ becomes:
\begin{equation}
\frac{P(w|\mathbf{y}_{u}^{k^\prime},\mathbf{y}_{p}^{\ell^\prime})
    \sum_{k} G(k^\prime)\sum_{\ell} W(\ell^\prime)
    }
    {
    \sum\limits_{k^{\prime\prime}}\sum\limits_{\ell^{\prime\prime}} P(w|\mathbf{y}_{u}^{k^{\prime\prime}},\mathbf{y}_{p}^{\ell^{\prime\prime}})
    \sum\limits_{k} G(k^{\prime\prime})
    \sum\limits_{\ell} W(\ell^{\prime\prime})},
\end{equation}
where 
\begin{itemize}
\setlength\itemsep{0.2em}
    \item[-] $G(\alpha) = P(\mathbf{y}_{u}^{\alpha}| \mathbf{z}_{u}^k)P(\mathbf{z}_{u}^k|u)$
    \item[-] $W(\beta) = P(\mathbf{y}_{p}^{\beta}| \mathbf{z}_{p}^\ell) P(\mathbf{z}_{p}^\ell|p)$
\end{itemize}

\subsection{M-step}
For the M-step, we assume the functional forms of the distributions for $P(\mathbf{z}_{u}^k|u)$, $P(\mathbf{z}_{p}^\ell|p)$ and $P(w|\mathbf{y}_{u}^{k^\prime},\mathbf{y}_{p}^{\ell^\prime})$ as:
\begin{align}
    P(\mathbf{z}_u|u) &\sim \text{Multinomial} \\
    P(\mathbf{z}_p|p) &\sim \text{Multinomial} \\
    P(w|u,p) &\sim \text{Multinomial}.
\end{align}

We introduce latent indicator variables
$\delta^{i,j}_{k,k^\prime,\ell,\ell^\prime}$
such that if the user $u^i$ belonging to latent user class $k$ (that was
corrupted through channel noise to latent user class $k^\prime$) used the word $w^i_j$ when reviewing
the product $p^i$ belonging to latent product class $\ell$ (corrupted by channel noise to latent product class $\ell^\prime$),
then
$\delta^{i,j}_{k,k^\prime,\ell,\ell^\prime}=1$, otherwise
$\delta^{i,j}_{k,k^\prime,\ell,\ell^\prime}=0$.

\noindent
The complete-data log likelihood is then:
\begin{equation}
\begin{split}
\mathcal{L}_C = &\sum_{i=1}^R
   \sum_{j=1}^{S_i} 
 \sum_{k^\prime=1}^K \sum_{\ell^\prime=1}^L 
\sum_{k=1}^K
\sum_{\ell=1}^L
\delta^{i,j}_{k,k^\prime,\ell,\ell^\prime} \\
& \bigg[ \log \bigg(P(w^i_j|\mathbf{y}_{u^i} = k^\prime, 
\mathbf{y}_{p^i} = \ell^\prime)  \big.  \\
& P(\mathbf{y}_{u^i} = k^{\prime}|\mathbf{z}_{u^i} = k) 
                        P(\mathbf{z}_{u^i} = k|u^i) \\
&P(\mathbf{y}_{p^i} = \ell^{\prime}|\mathbf{z}_{p^i} = \ell) P(\mathbf{z}_{p^i} =  \ell|p^i) \bigg) \bigg].
\end{split}    
\end{equation}
The expected complete data log-likelihood is:
\begin{equation}
\begin{split}
    \mathcal{L}_C  = &\sum^R_{i=1} \sum_{j=1}^{S_i} 
    \sum_{k^\prime=1}^K \sum_{\ell^\prime=1}^L
    \sum_{k=1}^K \sum_{\ell=1}^L
    E[\delta^{i,j}_{k,k^\prime,\ell,\ell^\prime}]\\
    &\left[ \log P(w_j^i|\mathbf{y}_{u^i}^{k^\prime},\mathbf{y}_{p^i}^{\ell^\prime}) + \log P(\mathbf{y}_{u^i}^{k^\prime}|\mathbf{z}_{u^i}^k) \right.\\
    &+ \left. \log P(\mathbf{y}_{p^i}^{\ell^\prime}|\mathbf{z}_{p^i}^\ell) + \log P(\mathbf{z}_{u^i}^k|u^i) + \log P(\mathbf{z}_{p^i}^\ell|p^i) \right].
\end{split}    
\end{equation}
For convenience, we will write $\mathcal{L}_C$ as:
\begin{equation}
\begin{split}
    \mathcal{L}_C  = &\sum^R_{i=1} \sum_{j=1}^{S_i} 
    \Delta^{u^i,p^i,w^i_j}_{u,p,w} \sum_{k^\prime=1}^K \sum_{\ell^\prime=1}^L
    \sum_{k=1}^K \sum_{\ell=1}^L
    E[\delta^{i,j}_{k,k^\prime,\ell,\ell^\prime}]\\
    &\left[ \log P(w|\mathbf{y}_{u}^{k^\prime},\mathbf{y}_{p}^{\ell^\prime}) + \log P(\mathbf{y}_{u}^{k^\prime}|\mathbf{z}_{u}^k) \right.\\
    &+ \left. \log P(\mathbf{y}_{p}^{\ell^\prime}|\mathbf{z}_{p}^\ell) + \log P(\mathbf{z}_{u}^k|u) + \log P(\mathbf{z}_{p}^\ell|p) \right],
\end{split}    
\end{equation}
where $\Delta^\alpha_a = 1$ if and only if $a = \alpha$, otherwise $\Delta^\alpha_a = 0$.
We maximize the expected complete data log-likelihood extended with Lagrange multiplier terms:
\begin{equation}
\begin{split}
    < \mathcal{L}_C > \quad = &\sum^R_{i=1} \sum_{j=1}^{S_i}
    \Delta^{u^i,p^i,w^i_j}_{u,p,w} \\
    &
    \sum_{k^\prime=1}^K \sum_{\ell^\prime=1}^L
    \sum_{k=1}^K \sum_{\ell=1}^L \hat P({\mathbf{z}}_{u}^k,\mathbf{{z}}_{p}^\ell, \mathbf{{y}}_{u}^{k^\prime},\mathbf{{y}}_{p}^{\ell^\prime}| u, p, w) \\
    &\left[ \log P(w|\mathbf{y}_{u}^{k^\prime},\mathbf{y}_{p}^{\ell^\prime}) + \log P(\mathbf{y}_{u}^{k^\prime}|\mathbf{z}_{u}^k) \right.\\
    &+ \left. \log P(\mathbf{y}_{p}^{\ell^\prime}|\mathbf{z}_{p}^\ell) + \log P(\mathbf{z}_{u}^k|u) + \log P(\mathbf{z}_{p}^\ell|p) \right] \\
    &+ \sum_{u \in \cal{U}} \lambda_u \left( \sum_k P(\mathbf{z}_{u}^k|u) - 1 \right) \\
    &+ \sum_{p \in \cal{P}} \lambda_p \left( \sum_\ell P(\mathbf{z}_{p}^\ell|p) - 1 \right) \\
    &+ \sum_{k^\prime,\ell^\prime} \lambda_{k^\prime,\ell^\prime} \left( \sum_{w \in \cal{V}} P(w|\mathbf{y}_{u}^{k^\prime}, \mathbf{y}_{p}^{\ell^\prime})  - 1 \right) .
\end{split}    
\end{equation}
By setting $\frac{\partial <\mathcal{L}_C>}{\partial P(w|\mathbf{y}_{u}^{k^\prime},\mathbf{y}_{p}^{\ell^\prime})} = 0$, we have:
\begin{equation}
\begin{split}
    &\frac{\partial <\mathcal{L}_C>}{\partial P(w|\mathbf{y}_{u}^{k^\prime},\mathbf{y}_{p}^{\ell^\prime})} = 0 \Longleftrightarrow \\
    & \sum_{k=1}^K \sum_{\ell=1}^L \hat P({\mathbf{z}}_{u}^k,\mathbf{{z}}_{p}^\ell, \mathbf{{y}}_{u}^{k^\prime},\mathbf{{y}}_{p}^{\ell^\prime}| u, p, w)
    \frac{1}{P(w|\mathbf{y}_{u}^{k^\prime},\mathbf{y}_{p}^{\ell^\prime})} \\
    &+ \lambda_{k^\prime,\ell^\prime} = 0.
\end{split}
\end{equation}
Then
\begin{equation}
\begin{split}
    P(w|\mathbf{y}_{u}^{k^\prime},\mathbf{y}_{p}^{\ell^\prime}) &=
    - \frac{\sum\limits_{(u,p) \in \mathcal{B}_w} \sum\limits_{k} \sum\limits_{\ell} \hat P({\mathbf{z}}_{u}^k,\mathbf{{z}}_{p}^\ell, \mathbf{{y}}_{u}^{k^\prime},\mathbf{{y}}_{p}^{\ell^\prime}| u, p, w)}
    {\lambda_{k^\prime,\ell^\prime}} \\
    &= - \frac{\sum\limits_{(u, p) \in \mathcal{B}_w} \hat P(\mathbf{y}_{u}^{k^\prime}, \mathbf{{y}}_{p}^{\ell^\prime}| u, p, w)}
    {\lambda_{k^\prime,\ell^\prime}},
\end{split}
\end{equation}
where ${\cal B}(w)$ is the set of (\textit{user, product}) tuples associated with the word $w$.
\noindent
Substituting $P(w|\mathbf{y}_{u}^{k^\prime},\mathbf{y}_{p}^{\ell^\prime})$ back to the constraint we have:
\begin{equation}
\begin{split}
    &\sum_{w} P(w|\mathbf{y}_{u}^{k^\prime},\mathbf{y}_{p}^{\ell^\prime}) = 1 \Longleftrightarrow \\
    &
    \lambda_{k^\prime,\ell^\prime} =
    - \sum_{w} \sum_{(u,p) \in {\cal B}(w)} \hat P(\mathbf{y}_{u}^{k^\prime}, \mathbf{{y}}_{p}^{\ell^\prime}| u, p, w),
\end{split}
\end{equation}
and so
\begin{equation}
\begin{split}
    P(w|\mathbf{y}_{u}^{k^\prime},\mathbf{y}_{p}^{\ell^\prime}) &= 
    \frac{\sum\limits_{(u,p) \in {\cal B}(w)} \hat P(\mathbf{{y}}_{u}^{k^\prime},\mathbf{{y}}_{p}^{\ell^\prime}| u, p, w)}
        {\sum\limits_{w^\prime} \sum\limits_{(u,p) \in {\cal B}(w^\prime)}  \hat P(\mathbf{{y}}_{u}^{k^\prime},\mathbf{{y}}_{p}^{\ell^\prime}| u, p, w^\prime)}.
\end{split}
\end{equation}
By setting $\frac{\partial <\mathcal{L}_C>}{\partial P(\mathbf{z}_{u}^k|u)} = 0$, we have:
\begin{equation}
\begin{split}
    &\frac{\partial <\mathcal{L}_C>}{\partial P(\mathbf{z}_{u}^k|u)} = 0 \Longleftrightarrow \\
    &\sum_{k^\prime=1}^K \sum_{\ell^\prime=1}^L \sum_{\ell=1}^L \hat P({\mathbf{z}}_{u}^k,\mathbf{{z}}_{p}^\ell, \mathbf{{y}}_{u}^{k^\prime},\mathbf{{y}}_{p}^{\ell^\prime}| u, p, w) \frac{1}{P(\mathbf{z}_{u}^k|u)} \\
    &+ \lambda_u = 0.
\end{split}
\end{equation}
Denoting the set of words used by user $u$ to review product $p$ by ${\cal W}(u,p)$, we obtain:
\begin{equation}
\begin{split}
    P(\mathbf{z}_{u}^k|u) &=
    - \frac{\sum\limits_p \sum\limits_{w \in {\cal W}(u,p)} \sum\limits_{k^\prime} \sum\limits_{\ell^\prime} \sum\limits_{\ell} \hat P({\mathbf{z}}_{u}^k,\mathbf{{z}}_{p}^\ell, \mathbf{{y}}_{u}^{k^\prime},\mathbf{{y}}_{p}^{\ell^\prime}| u, p, w)}
    {\lambda_u} \\
    &= - \frac{\sum\limits_p \sum\limits_{w \in{\cal W}(u,p)} \sum\limits_\ell \hat P(\mathbf{{z}}_{u}^k,\mathbf{{z}}_{p}^{\ell}| u, p, w)}{\lambda_u}.
\end{split}
\end{equation}
Substituting $P(\mathbf{z}_{u}^k|u)$ back to the constraint we have:
\begin{equation}
\begin{split}
    &\sum_{k=1}^K P(\mathbf{z}_{u}^k|u) = 1 \\
    & \Longleftrightarrow \lambda_u = - \sum_k \sum_\ell \sum_p \sum_{w \in{\cal W}(u,p)} \hat P(\mathbf{{z}}_{u}^k,\mathbf{{z}}_{p}^{\ell}| u, p, w) \\
    & \Longleftrightarrow \lambda_u = - \sum_p |{\cal W}(u,p)|
\end{split}
\end{equation}
and so
\begin{equation}
\begin{split}
    P(\mathbf{z}_{u}^k|u) &= 
    \frac{\sum_p \sum_{w \in{\cal W}(u,p)} \sum_\ell \hat P(\mathbf{{z}}_{u}^k,\mathbf{{z}}_{p}^{\ell}| u, p, w)}{\sum_p |{\cal W}(u,p)|}.
\end{split}
\end{equation}
By setting $\frac{\partial <\mathcal{L}_C>}{\partial P(\mathbf{z}_{p}^\ell|p)} = 0$, we have:
\begin{equation}
\begin{split}
    &\frac{\partial <\mathcal{L}_C>}{\partial P(\mathbf{z}_{p}^\ell|p)}  = 0 \Longleftrightarrow \\
    &\sum_{k^\prime=1}^K \sum_{\ell^\prime=1}^L \sum_{k=1}^K  \hat P({\mathbf{z}}_{u}^k,\mathbf{{z}}_{p}^\ell, \mathbf{{y}}_{u}^{k^\prime},\mathbf{{y}}_{p}^{\ell^\prime}| u, p, w) \frac{1}{P(\mathbf{z}_{p}^\ell|p)} + \lambda_p = 0.
\end{split}
\end{equation}
Then,
\begin{equation}
\begin{split}
    P(\mathbf{z}_{p}^\ell|p) &=
    - \frac{\sum\limits_u \sum\limits_{w \in{\cal W}(u,p)} \sum\limits_{k^\prime} \sum\limits_{\ell^\prime} \sum\limits_{k} \hat P({\mathbf{z}}_{u}^k,\mathbf{{z}}_{p}^\ell, \mathbf{{y}}_{u}^{k^\prime},\mathbf{{y}}_{p}^{\ell^\prime}| u, p, w)}{\lambda_p} \\
    &= - \frac{\sum\limits_u \sum\limits_{w \in{\cal W}(u,p)} \sum\limits_k \hat P(\mathbf{{z}}_{u}^k,\mathbf{{z}}_{p}^{\ell}| u, p, w)}{\lambda_p}.
\end{split}
\end{equation}
Substituting $P(\mathbf{z}_{p}^\ell|p)$ back to the constraint we have:
\begin{equation}
\begin{split}
     &\sum_\ell P(\mathbf{z}_{p}^\ell|p) = 1 \\
    &\Longleftrightarrow \lambda_p =-   \sum_k \sum_\ell \sum_u \sum_{w \in{\cal W}(u,p)} \hat P(\mathbf{{z}}_{u}^k,\mathbf{{z}}_{p}^{\ell}| u, p, w) \\
    & \Longleftrightarrow \lambda_p = - \sum_u |{\cal W}(u,p)|
\end{split}
\end{equation}
and finally:
\begin{equation}
\begin{split}
    P(\mathbf{z}_{p}^\ell|p) &= 
    \frac{\sum_u \sum_{w \in{\cal W}(u,p)} \sum_k 
    \hat P(\mathbf{{z}}_{u}^k,\mathbf{{z}}_{p}^{\ell}| u, p, w)}{\sum_u |{\cal W}(u,p)|}.
\end{split}
\end{equation}

\end{document}